\documentclass[useAMS,usenatbib]{mnras}
\usepackage[dvips]{graphicx}
\usepackage{amsmath}
\usepackage{natbib}
\usepackage{pifont}
\usepackage{txfonts}
\usepackage{mathrsfs}
\usepackage{amsfonts}
\usepackage{amssymb}
\usepackage{hyperref}
\usepackage{epsfig}
\usepackage{threeparttable}

\title[The disappearing act: RW Aur]{The disappearing act: A dusty wind eclipsing RW Aur}

\author[I. Bozhinova et al.]{
      I. Bozhinova$^{1}$, A. Scholz$^{1}$, G. Costigan$^{2}$, O. Lux$^{3}$, C. J. Davis$^{4}$,  T. Ray$^{5}$, N. F. Boardman$^{1}$, \newauthor  K. L. Hay$^{1}$,  T. Hewlett$^{1}$, G. Hodos\'an$^{1}$, B. Morton$^{1}$  \\ \\
      $^{1}$SUPA, School of Physics and Astronomy, University of St Andrews, North Haugh, St Andrews,  KY16 9SS, UK \\
      $^{2}$Leiden Observatory, University of Leiden, PB 9513, NL-2300 RA Leiden, the Netherlands \\
	  $^{3}$Astrophysical Institute and University Observatory, Schillerg\"asschen 2-3, 07745 Jena, Germany \\     
      $^{4}$Liverpool John Moores University, Astrophysics Research Institute, Liverpool Science Park, IC2 Building
146 Brownlow Hill, Liverpool, L3 5RF, UK \\ 
      $^{5}$Dublin Institute for Advanced Studies, 31 Fitzwilliam Place, Dublin 2, Ireland
    }

\begin{document}    
    
\maketitle

  \begin{abstract}
      
	RW Aur is a young binary star that experienced a deep dimming in 2010-11 in component A and a second even deeper dimming from summer 2014 to summer 2016. We present new unresolved multi-band photometry during the 2014-16 eclipse, new emission line spectroscopy before and during the dimming, archive infrared photometry between 2014-15, as well as an overview of literature data.
	
	Spectral observations were carried out with the Fibre-fed RObotic Dual-beam Optical Spectrograph on the Liverpool Telescope. Photometric monitoring was done with the Las Cumbres Observatory Global Telescope Network and James Gregory Telescope. Our photometry shows that RW Aur dropped in brightness to R = 12.5 in March 2016. In addition to the long-term dimming trend, RW Aur is variable on time scales as short as hours. The short-term variation is most likely due to an unstable accretion flow. This, combined with the presence of accretion-related emission lines in the spectra suggest that accretion flows in the binary system are at least partially visible during the eclipse.  
	 
The equivalent width of [O\,I] increases by a factor of ten in 2014, coinciding with the dimming event, confirming previous reports. The blue-shifted part of the $H\alpha$ profile is suppressed during the eclipse. In combination with the increase in mid-infrared brightness during the eclipse reported in the literature and seen in WISE archival data, and constraints on the geometry of the disk around RW Aur A we arrive at the conclusion that the obscuring screen is part of a wind emanating from the inner disk.

  \end{abstract}
   

\section{Introduction}

Classical T Tauri stars (TTS) are pre-main-sequence objects with strong irregular variability, first noted in a study of highly variable stars by \cite{1945ApJ...102..168J}. In that same work, RW Aur is identified as a T Tau type star with a spectrum rich in emission lines. Today, it is known that RW Aur is a visual binary with a separation of 1.4\arcsec \,between the primary and secondary components. The spectrum of the primary shows strong signatures of accretion and winds (eg. \citealt{2001A&A...369..993P,2005A&A...440..595A}). The secondary, RW Aur B, is a late K type TTS \citep{1993AJ....106.2005G,1997ApJ...490..353G,2004ApJ...616..998W} with little or no evidence of ongoing accretion. There are indications for the existence of more stellar/substellar companions in the system. The multiplicity study of T Tau stars by \cite{1993AJ....106.2005G} reveals a possible tertiary component (RW Aur C) orbiting RW Aur B, with a separation of 0.12\arcsec\,between the B and C stars. Furthermore, spectroscopic studies by \cite{1999A&A...352L..95G} and \cite{2001A&A...369..993P} reveal periodic variations in the radial velocity of a number of lines in RW Aur A. The authors explore the possibility of a brown dwarf sized companion (RW Aur D) in close orbit around the primary as the cause for the observed spectroscopic variations. 

\cite{2006A&A...452..897C} examined CO maps of RW Aur from interferometry. They estimated the size of the circumstellar disk around RW Aur A to be unusually small (40-57\,AU) and found evidence for a 600\,AU long structure of material trailing from the primary star. The disk size was calculated for a disk inclination between 45-60$^{\circ}$. Both the size of the disk and the $``$arm$"$  were suggested as evidence for the tidal disruption of the circumstellar disk from a recent interaction with the secondary RW Aur B. This tidal interaction hyphothesis is further backed up by recent hydrodynamical simulations of the RW Aur disk by \cite{2015MNRAS.449.1996D}. 
   
In 2010 RW Aur underwent a long lasting dimming event, that had never been observed before. The star dropped by $\sim$2 mag in the optical and remained in that state for several months. \cite{2013AJ....146..112R} presented a photometric investigation of this event, looking at KELT and American Association of Variable Star Observers (AAVSO) data. They conclude that the most likely explanation of this event is obscuration by material located $\sim$180\,AU from RW Aur A. It is noted that the estimated occulting body lies far beyond the outer edge of the disk, but is still within the size of the tidally disrupted arm and therefore likely a part of it. 

\begin{figure}
	\includegraphics[width=1.0\linewidth]{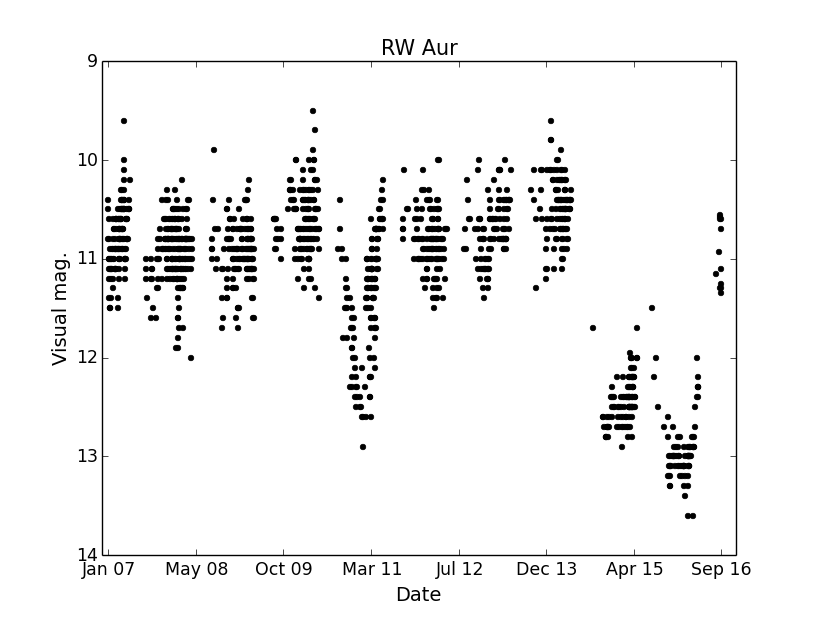}
	\caption{American Association of Variable Star Observers (AAVSO) visual lightcurve for RW Aur between 2007 and 2016. The 2010 and 2014 dimmings are clearly visible.}
	\label{aavso}
\end{figure}

In 2014 RW Aur entered a second long lasting, even deeper ($\sim$3 mag, \cite{2015IBVS.6126....1A}) minimum. Figure \ref{aavso} shows an AAVSO visual lightcurve of RW Aur up until August 2016, including both dimming events. As can be seen in this lightcurve, there are clear signs that the star has recovered to its out-of-eclipse brightness
in August 2016, indicating the end of the second eclipse. Our own photometry on Aug 25 and 26 confirms this,
with R-band magnitudes of 10.7 and 10.6 \citep{2016ATel.9428....1S}.
 
\cite{2015A&A...577A..73P} performed a spectroscopic analysis of emission lines associated with accretion, wind and jets in and out of the 2014 minimum. They found enhancement in the equivalent widths of [O\,I] and [S\,II] (forbidden lines associated with jets), but no increase in the fluxes. Additionally, an increase in the strength of the resonant lines Ca\,II and Na\,I was reported. Furthermore, they found no change in the emission of H$\alpha$ and He\,I, lines associated with accretion activity. The authors concluded that the obscuring body only covers the star and inner regions of the system, while the outer parts of the wind and jets remained unaffected. This effect was  attributed to dust grains being lifted up into the line of sight through interactions with the winds. In addition, \cite{2015IBVS.6143....1S} report results of infrared photometry in JHKML in the period 2010-2015. They find a drop in JHK brightness corresponding to the 2014 dimming event, but a simultaneous increase in brightness in M and L. The excess IR emission is attributed to hot dust of about 1000\,K, located around the inner disk rim, 0.1-0.2\,AU away from the star. It is further speculated that the hot dust could then be lifted up into the wind, supporting the scenario presented in \cite{2015A&A...577A..73P}.

The dusty wind idea is in contrast to the tidal arm dimming scenario \citep{2013AJ....146..112R,2016AJ....151...29R} where an occultation by the tidal arm is expected to at least partially obscure the inner disk and possibly cover the excess IR emission. \cite{2016AJ....151...29R} also report a smaller dimming in 2012-13 and argue that the kinematics of this event fit the values derived from their analysis of the 2010 eclipse. Their analysis of the 2014 dimming using the same method is hampered by the fact that the ingress for the eclipse happened during the seasonal observing gap and its duration is not constrained yet. The tidal arm dimming scenario is potentially supported by the hydrodynamical simulations by \cite{2015MNRAS.449.1996D}. Their simulations suggest the presence of material in a bridge between RW Aur A and B, with several particles also crossing the line-of-sight to the system.

Finally, \cite{2016ApJ...820..139T} present a spectroscopic study of the system between Oct 2010 and Jan 2015 and argue that neither the tidal arm nor dusty wind scenario alone can account for the complex variability seen in the lines in and out of the eclipses. The authors further explore time-variable mass accretion as an alternative explanation to the observed dim and bright states of the system within the 2010-15 time frame.  

In this study we present photometric and spectroscopic observations of RW Aur taken just before and during the 2014 dimming. We explore the photometric variability in several optical and infrared bands along with equivalent width and profile analysis of the  H$\alpha$, He\,I and [O\,I] lines to provide further evidence to the above scenarios.

\section{Observations and Data Reduction}
\subsection{LCOGT}
A portion of the photometric data presented in this work was taken with the Las Cumbres Observatory Global Telescope Network (LCOGT) \citep{2013PASP..125.1031B} as part of programs STA2014B-002 and STA2015A-002. This set spans the period 26 Dec 2014 - 11 Jan 2016. All images were taken with SBIG cameras on 1-meter telescopes. The detectors have a size of 4K\,$\times$\,4K pixels, 9$\,\mu m$ pixel size, field of view of 15.8\,$\times$\,15.8\,\arcmin and resolution of 0\,\farcs464\,\slash pix. Each epoch consist of a set of images of 10s and 100s exposure time in four bandpasses (Sloan {\it u\,\arcmin, g\,\arcmin, r\,\arcmin} and {\it i\,\arcmin}, \citealt{1996AJ....111.1748F}). In some cases, where observations have been interrupted and resumed, we have multiple images per epoch. The main science frames are the 100s exposures, the shorter exposures were taken in the event of saturation of objects in the field, but have not been used in the current work. Table \ref{LCOGTobs} summarises the epochs and exposures for each filter. The raw images have been reduced with the LCOGT data reduction pipeline\footnote{http://lcogt.net/observatory/data/pipeline/}.     

\begin{table}
	\caption{Summary of epochs and number of 100s exposures per epoch in each filter from the LCOGT data set.}
	\label{LCOGTobs}
	\centering
	\begin{tabular}{|c|c|c|c|c|c|}
		\hline 
 		{MJD} & {\bf Date} & {\bf u\,\arcmin } & {\bf g\,\arcmin } & {\bf r\,\arcmin} & {\bf i\,\arcmin } \\ 
		&&  	(3543\,$\AA$)  & (4770\,$\AA$) &  (6231\,$\AA$) & (7625\,$\AA$) \\
 		\hline 
		57017 & 26-12-2014 & 0 & 3 & 3 & 3 \\ 
		57020 & 29-12-2014 & 1 & 2 & 2 & 2 \\ 
		57038 & 16-01-2015 & 1 & 1 & 1 & 1 \\  
		57041 & 19-01-2015 & 1 & 1 & 1 & 2 \\ 
		57043 & 21-01-2015 & 1 & 1 & 1 & 1 \\ 
		57046 & 24-01-2015 & 1 & 1 & 1 & 1 \\ 
		57049 & 27-01-2015 & 1 & 2 & 2 & 2 \\ 
		57051 & 29-01-2015 & 1 & 3 & 2 & 2 \\ 
		57056 & 03-02-2015 & 1 & 1 & 1 & 1 \\ 
		57058 & 05-02-2015 & 1 & 1 & 1 & 1 \\ 
		57062 & 09-02-2015 & 0 & 2 & 2 & 2 \\  
		57077 & 24-02-2015 & 1 & 1 & 1 & 1 \\  
		57088 & 07-03-2015 & 1 & 1 & 1 & 1 \\ 
		57092 & 11-03-2015 & 1 & 1 & 1 & 2 \\ 
		57097 & 16-03-2015 & 1 & 0 & 1 & 1 \\ 
		57104 & 23-03-2015 & 1 & 1 & 1 & 1 \\ 
		57109 & 28-03-2015 & 0 & 1 & 1 & 1 \\ 
		57258 & 24-08-2015 & 1 & 1 & 1 & 1 \\
		57260 & 26-08-2015 & 1 & 1 & 1 & 1 \\		
		57262 & 28-08-2015 & 1 & 1 & 1 & 1 \\
		57264 & 30-08-2015 & 1 & 1 & 1 & 1 \\
		57266 & 01-09-2015 & 1 & 1 & 1 & 1 \\
		57268 & 03-09-2015 & 1 & 1 & 1 & 1 \\
		57270 & 05-09-2015 & 1 & 1 & 1 & 1 \\
		57278 & 13-09-2015 & 1 & 1 & 1 & 1 \\
		57280 & 15-09-2015 & 1 & 1 & 1 & 1 \\
		57292 & 27-09-2015 & 1 & 1 & 1 & 1 \\
		57294 & 29-09-2015 & 1 & 1 & 1 & 1 \\
		57315 & 20-10-2015 & 1 & 1 & 1 & 1 \\
		57321 & 26-10-2015 & 1 & 1 & 1 & 1 \\
		57323 & 28-10-2015 & 2 & 2 & 2 & 1 \\
		57352 & 26-11-2015 & 1 & 1 & 1 & 1 \\
		57360 & 04-12-2015 & 1 & 1 & 1 & 1 \\
		57365 & 09-12-2015 & 1 & 1 & 1 & 1 \\
		57367 & 11-12-2015 & 1 & 1 & 1 & 1 \\
		57372 & 16-12-2015 & 1 & 1 & 1 & 1 \\
		57375 & 19-12-2015 & 1 & 1 & 1 & 1\\
		57377 & 21-12-2015 & 1 & 1 & 1 & 1\\
		57380 & 24-12-2015 & 1 & 1 & 1 & 1 \\
		57382 & 26-12-2015 & 1 & 1 & 1 & 1 \\
		57395 & 08-01-2016 & 1 & 1 & 1 & 1\\
		57398 & 11-01-2016 & 1 & 1 & 1 & 1 \\
		\hline
	\end{tabular}
\end{table} 

\subsection{JGT}
As a complement to the LCOGT observations, we have also monitored RW Aur with the 0.94-meter James Gregory Telescope (JGT) at the University of St Andrews Observatory. The JGT has an Andor CCD detector, 1K$\,\times\,$1K pixels, 13$\mu m$ pixel size, resolution of $\sim$ 1\,\farcs0\,\slash pix and a field of view of about 15\arcmin$\,\times\,$15\arcmin. At each epoch we continuously monitored the target for a period of at least 1 hour in the Cousin R-band in order to explore its short time-scale behaviour. The exposure time is 30-60\,s, depending on conditions. The JGT dataset (summarised in Table \ref{JGTobs}) was taken between March 2015 and March 2016. Standard reduction including bias subtraction and flat-fielding was performed. 

\begin{table}
	\caption{Summary of photometric observations with the JGT. This table lists date, number of images, exposure time and average Cousin R-band magnitude of RW Aur for each night.}
	\label{JGTobs}
	\centering
	\begin{tabular}{|c|c|c|c|c|}
		\hline 
		MJD & Date & no images & t exp. (s) & R (mag) \\ \hline 
		57092 & 11-03-2015 & 270 & 60 & 11.89 \\ 
		57099 & 18-03-2015 & 160 & 60 & 12.21 \\ 
		57104 & 23-03-2015 & 110 & 60 & 11.99 \\ 
		57110 & 29-03-2015 & 130 & 60 & 12.08 \\ 
		57337 & 11-11-2015 & 240 & 30 & 12.71 \\		
		57367 & 11-12-2015 & 210 & 60 & 12.47 \\
		57368 & 12-12-2015 & 270 & 30 & 12.53 \\
		57401 & 14-01-2016 & 250 & 60 & 12.54 \\
		57428 & 10-02-2016 & 320 & 60 & 12.65 \\
		57429 & 11-02-2016 & 240 & 60 & 12.61 \\
		57454 & 07-03-2016 & 240 & 30 & 12.41 \\
		57478 & 31-03-2016 & 110 & 60 & 12.43 \\

		\hline 
	\end{tabular} 
\end{table}

\subsection{Spectral observations} 

Spectral observations were performed with FRODOSpec (Fibre-fed RObotic Dual-beam Optical Spectrograph), a robotic integral field spectrograph on the Liverpool Telescope \citep{2004SPIE.5489..679S}. Data covers 19 nights between February 2013 and February 2015 with three spectra taken each night (Table \ref{EWstable}). Both blue (3900\,-\,5100\AA) and red arm (5900\,-\,8000\,\AA) observations were completed. Due to the very low S/N, the blue arm spectra were not considered in the following work. The resolution of the red arm in high resolution mode is $\sim$ 5,300. The integral field covers 9\,\farcs84$\times\,$9\,\farcs84, with 12 lenslets each with diameter of 0\,\farcs82 on-sky. The exposure time is 300s for all spectra.

Due to the robotic nature of the telescope, in some of the observations the source is either not within the integral field unit (IFU) or is not centred within the IFU. The main result appears to be an increase in the noise within the spectrum. We only considered the cases where the PSF peak lies within the IFU. In the majority of cases the full PSF lies completely within the IFU. In some cases (see Table \ref{EWstable}), the peak of the PSF lies on the edge of the detector (as noted by the spread in the PSF across the IFU edge), and so we are missing flux from one of the wings of the PSF. We tested for differences in the measured EW in these epochs and no significant difference was seen. 

We reduce the data using the FRODOSpec pipeline \citep{2012AN....333..101B}. The pipeline first performs basic CCD processing, including bias subtraction, overscan trimming and CCD flat fielding. A continuum lamp is used to find the fibre positions and find a polynomial trace for each. Flux is extracted for each fibre in the target frame, continuum and arc lamp frames. Wavelength calibration is then performed for each spectrum and transmission correction, using the continuum lamp. The spectra are rebinned in order to determine a single wavelength solution for the entire data cube. Finally, sky extraction is done by combining the signal from sky only fibers. The spectra are not flux calibrated. We also do not resolve RW Aur A and B. 

\begin{table}
\caption{Summary of spectral observation epochs and mean equivalent width measurements for the H$\alpha$, [O\,I] and He\,I emission Line. All EWs are in $\AA$. Epochs in which the target was not centred in the IFU i.e. not completely covered, are indicated in the last column.}
\label{EWstable}
\begin{tabular}{cccccc}
\hline
MJD & Date & H\,$\alpha$ &  [O\,I]  &  He\,I  & IFU coverage \\
\hline \hline
56330 & 07-02-2013 & 83.88 &      0.54 &      1.01 &  Partial \\
56363 & 12-13-2013 &  88.07 &      0.54 &     0.37 &   Complete\\
56500 & 27-07-2013 &  77.94 &      0.7 &      0.42 &   Complete \\
56524 & 20-08-2013 &  74.87 &      0.62 &      1.09 &     Complete\\
56544 & 09-09-2013 &  55.39 &      0.37 &     0.6 &    Complete\\
56565 & 30-09-2013 &  35.53 &      0.36 &      0.49 &    Complete\\
56586 & 21-10-2013 &  60.79 &      0.55 &      0.64 &    Complete\\
56611 & 15-11-2013 &  78.96 &      0.49 &      0.64 &    Complete\\
56644 & 18-12-2013 &  42.68 &      0.69 &    1.11 &    Complete\\
56671 & 14-01-2014 &  51.9 &      0.61 &      0.65 &  Complete\\
56724 & 08-03-2014 &  52.21 &      0.48 &      0.35 &   Complete\\
56740 & 24-03-2014 &  45.39 &      0.55 &      0.43 &    Complete\\
56752 & 05-04-2014 &  69.7 &      0.54 &      1.13 &  Complete\\
56910 & 10-09-2014 &  21.29 &      1.9 &      0.5 &  Partial \\
57028 & 06-01-2015 &  17.97 &      2.61 &      0.4 &  Partial  \\
57034 & 12-01-2015 &  16.05 &      4.63 &      0.14 &  Complete\\
57039 & 17-01-2015 &  32.25 &      4.96 &      0.32 &  Partial \\
57049 & 27-01-2015 &  26.59 &      5.75 &     0.18 &   Partial  \\
57079 & 26-02-2015 &  23.81 &      3.64 &      0.72 &  Partial\\
\hline
\end{tabular}
\end{table}
	
\section{Data analysis}	
\label{data_red}

\subsection{Photometry}

For the LCOGT and JGT images we perform aperture photometry using the standard photometric IRAF package {\it daophot}\,\footnote{http://iraf.noao.edu/docs/photom.html}. We measure RW Aur along with 10-15 well exposed, unsaturated comparison stars.
The aperture is set to 2.5 times the FWHM of the PSF. We construct differential lightcurves 
by calculating a reference lightcurve from well-exposed comparison stars and subtracting that from the RW Aur
raw lightcurve. This ensures systematic effects such as variable weather conditions are removed from the final data.
The reference curves were inspected for any outlying points and a few epochs were removed (due to proximity of the Moon). 

\begin{figure}
	\begin{tabular}{c}
		\includegraphics[width=0.94\linewidth]{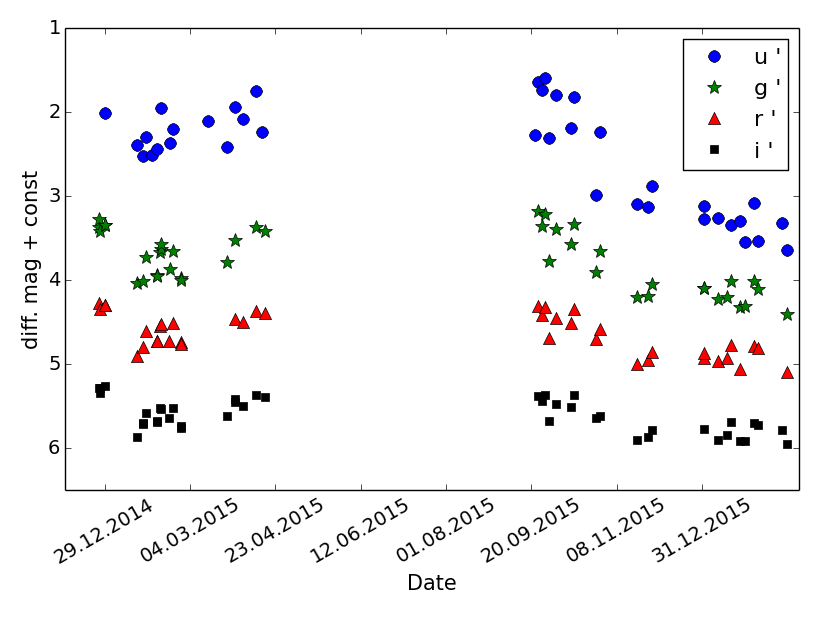} \\
	\end{tabular}
	
	\caption{
	RW Aur lightcurves from LCOGT data. {\it Blue circles} - u\,\arcmin\, band (3543$\,\AA$), 
		{\it green stars} - g\,\arcmin\, band (4770$\,\AA$), {\it red triangles} - r\,\arcmin\, band (6231$\,\AA$),
		{\it black squares} - i\,\arcmin\, band (7625$\,\AA$). The lightcurves are shifted by an arbitrary constant for clarity. All lightcurves show a similar shape.
		}
	\label{LCOGTlc}
\end{figure}

The lightcurves derived from the LCOGT data are presented in Figure \ref{LCOGTlc}. 
Photometric accuracy (defined as the lower envelope of the RMS of the reference stars) is on the order of a few percent, with RW Aur being a clear, highly variable outlier. The RW Aur curves in the four bands have similar shapes and amplitudes (between 0.7-2.1\,mag, Table \ref{Aslopes}). The amplitudes are calculated as the max-min difference with an uncertainty of about 0.02 mag. A visual inspection shows a very good correlation between the shape of the lightcurves in the r\,\arcmin and i\,\arcmin-bands, but the correlation becomes weaker towards the bluer wavelengths. The brightness of RW Aur has continued to decline throughout the 2015-2016 winter observing season in all filters. The similarity in the shapes of the lightcurves in the separate bands before and after the 2015 observing gap suggests the continued drop in flux is caused by the same source as the initial eclipse in 2014.

\begin{table}
	\centering
	\caption{Amplitudes (A) and reddening slopes for the observed data and interstellar extinction. The error for the observed amplitudes is 0.02\,mag. The observed slopes correspond to the best (linear regression) fit to the photometric data.
	We use the extinction law by \citealt{1989ApJ...345..245C} to calculate extinction amplitudes and reddening slopes. In this table, we present the amplitudes and slopes calculated for $R_V=3.1$ and $A_V=0.9$, which are the parameter values producing the best fit to the observed r-band amplitude.
	}
	\label{Aslopes}	
	\begin{tabular}{|c|c|c|}
		\hline 
		{\bf Bands} & {\bf Obs. A} & {\bf Exct. A} \\ \hline
		u & 2.05 & 1.49\\ 
		g & 1.22 & 1.12 \\ 
		r & 0.81 & 0.81 \\ 
		i & 0.70 & 0.61 \\ 
		\hline 
		{\bf Color} & {\bf Obs. slope} & {\bf Exct. slope} \\ \hline
		u-g & 1.82  & 4.08  \\ 		
		g-r & 2.8 & 3.68  \\ 
		r-i & 3.89  & 4.02 \\ 
		\hline 
	\end{tabular}
\end{table} 

\begin{figure*}
	\begin{tabular}{cc}
		\includegraphics[width=0.5\linewidth]{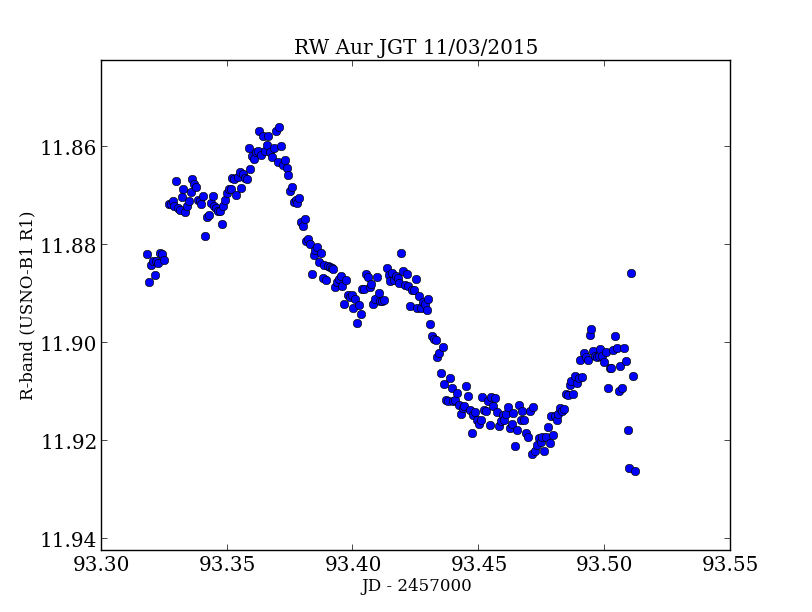}
		\includegraphics[width=0.5\linewidth]{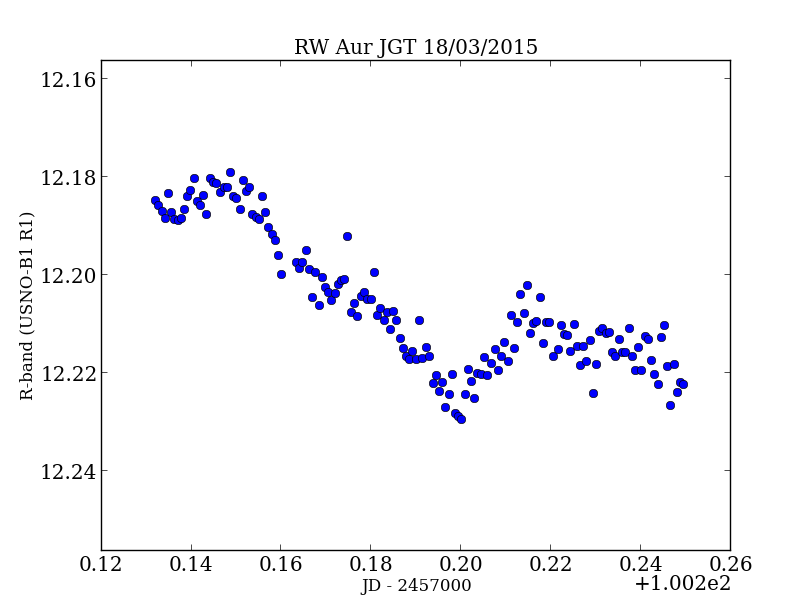} \\
		\includegraphics[width=0.5\linewidth]{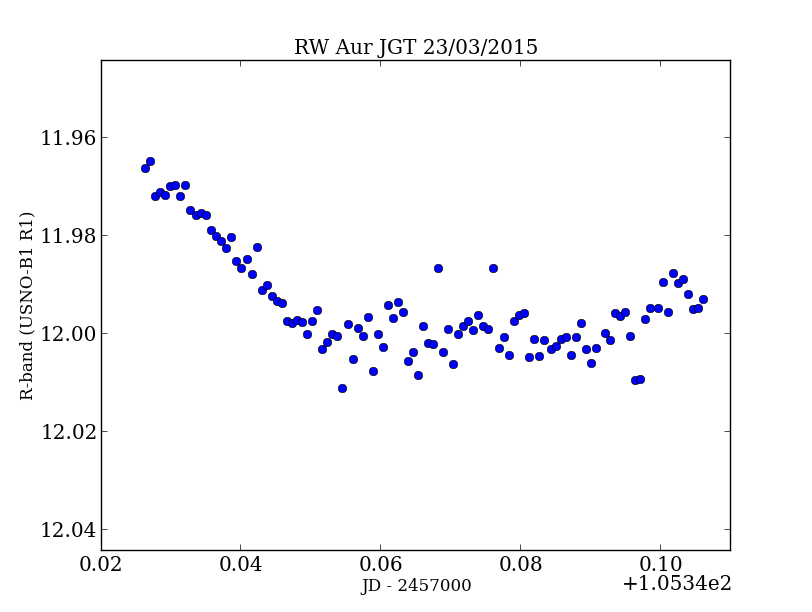} 
		\includegraphics[width=0.5\linewidth]{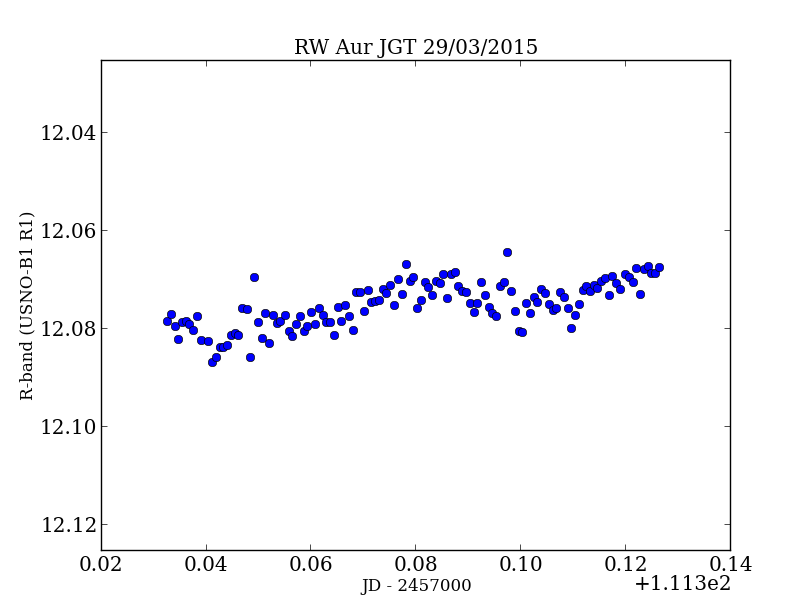} \\
		\includegraphics[width=0.5\linewidth]{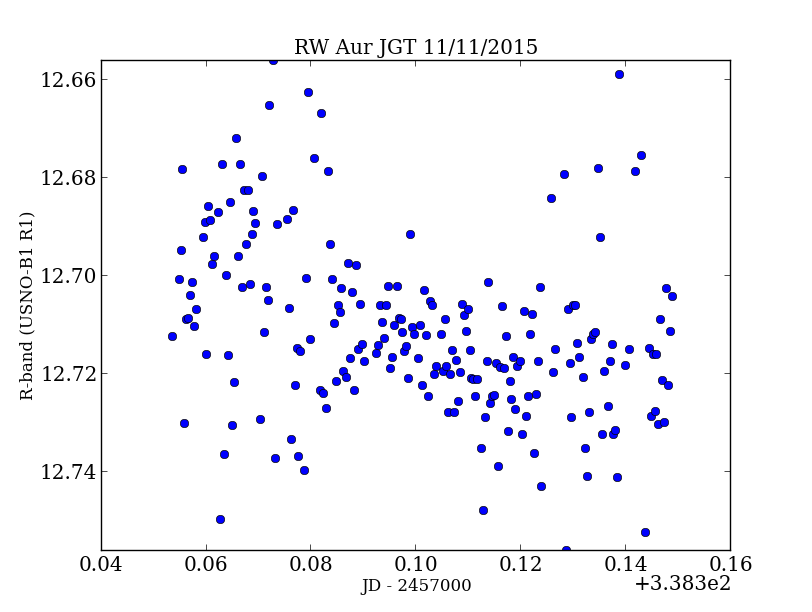} 
		\includegraphics[width=0.5\linewidth]{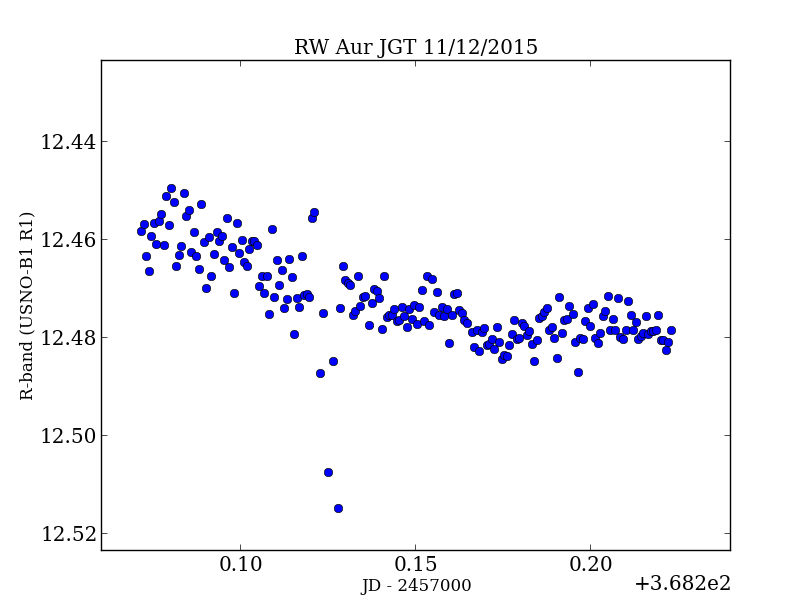} \\
	\end{tabular}
	\caption{RW Aur JGT lightcurves, R-band. Significant irregular variability (up to 0.2 mag) on time-scales of hours can be seen throughout the entire observing season.}
	\label{JGTlc}
\end{figure*}

Standard aperture photometry was applied to the JGT images using $\sim$6 reference stars. 
The R-band was shifted to standard using the USNO-B1.0 magnitudes for 3 reference stars
\citep{2003AJ....125..984M}. The error is $\sim$0.01 mag. The differential lightcurves are presented in Figure \ref{JGTlc}. The lightcurves show a significant irregular variability (up to 0.2 mag) on timescales of hours. 

\renewcommand{\thefigure}{\arabic{figure} (continued)}
\addtocounter{figure}{-1}
 
\begin{figure*}
	\begin{tabular}{cc}
		\includegraphics[width=0.5\linewidth]{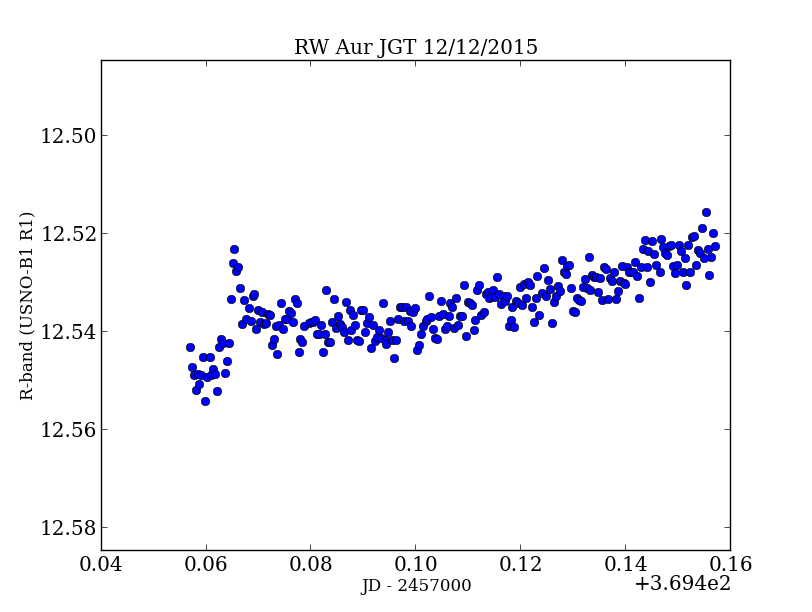}
		\includegraphics[width=0.5\linewidth]{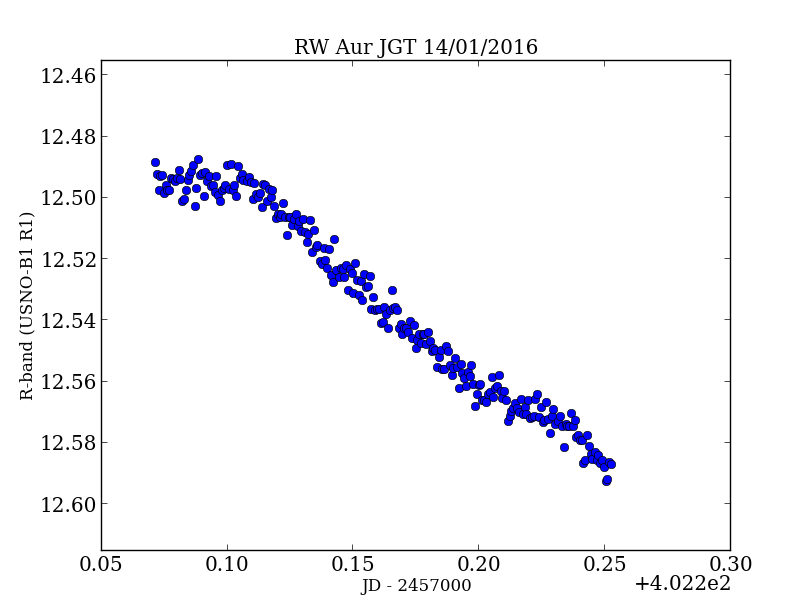} \\
		\includegraphics[width=0.5\linewidth]{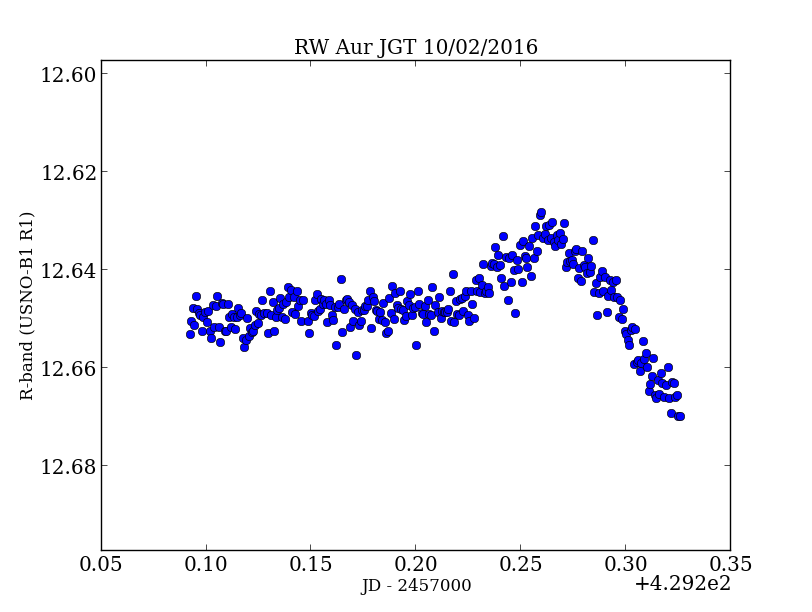} 
		\includegraphics[width=0.5\linewidth]{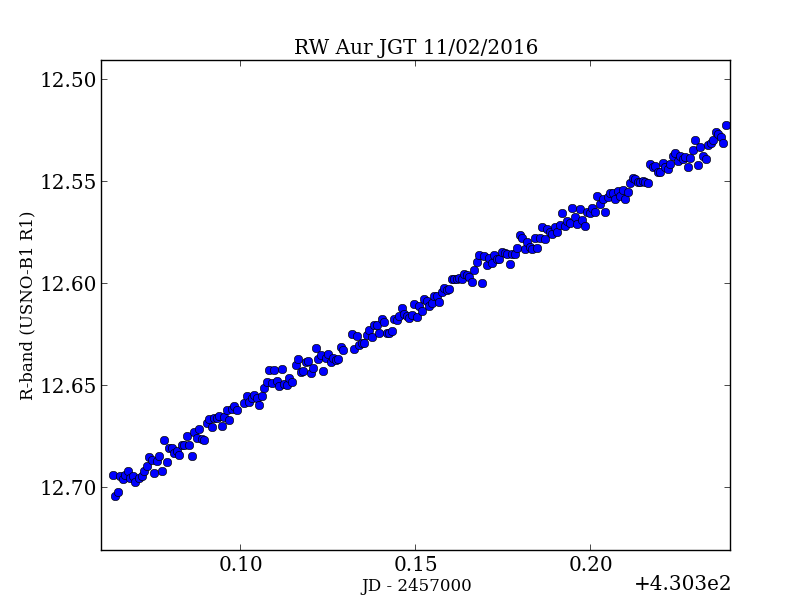} \\
		\includegraphics[width=0.5\linewidth]{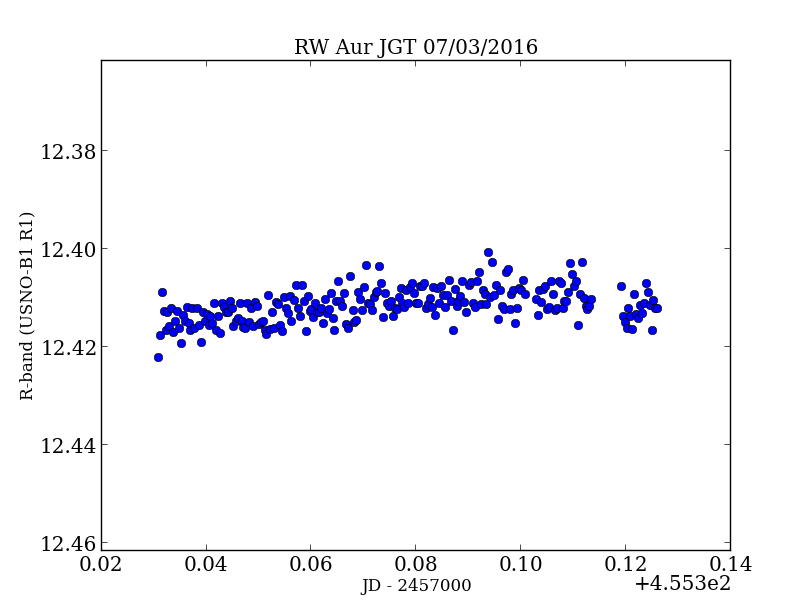} 
		\includegraphics[width=0.5\linewidth]{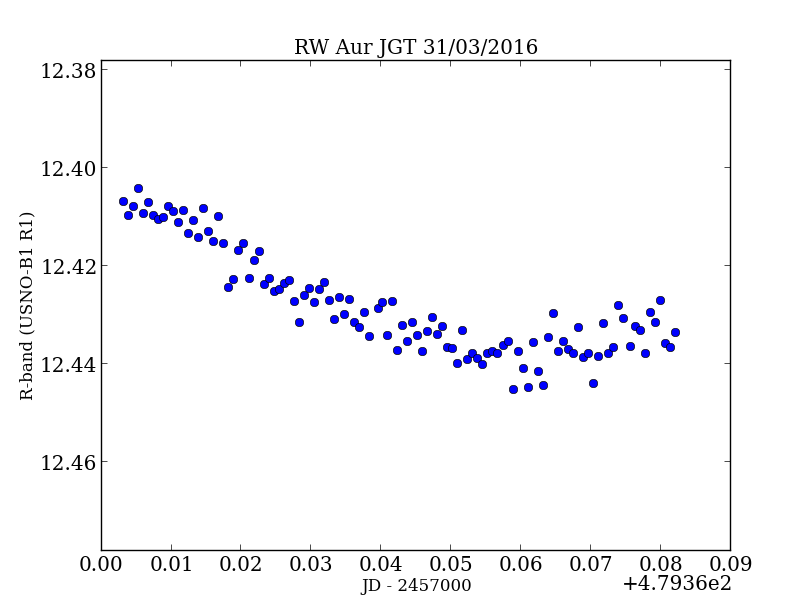} \\
	\end{tabular}
	\caption{}
\end{figure*}
 
\renewcommand{\thefigure}{\arabic{figure}}

\subsection{Mid-IR WISE archival data}
\label{neowisesec}

\begin{figure}
	\includegraphics[width=1.0\linewidth]{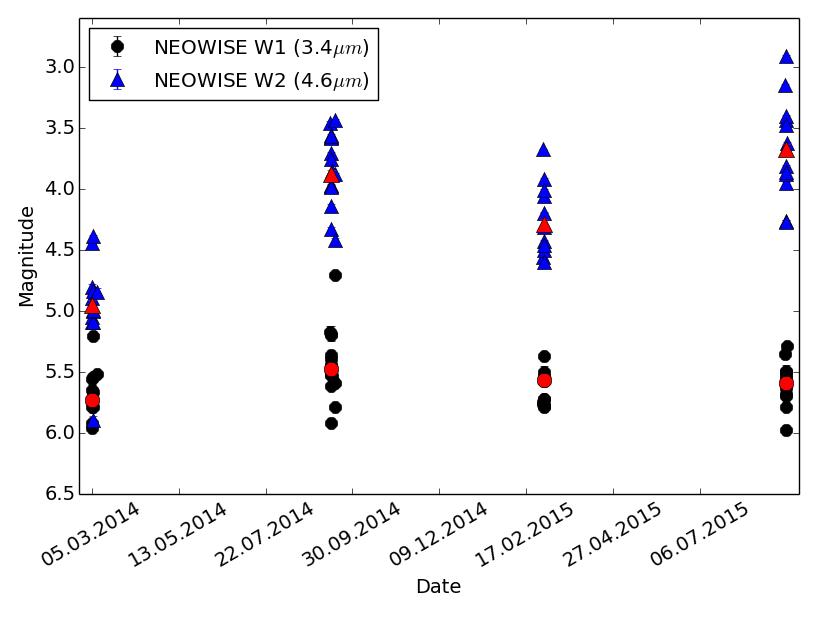}
	\caption{NEOWISE observations of RW Aur before and after the 2014 dimming. The red datapoints represent the average magnitude of RW Aur for each observational block. The system becomes brighter in both bands during the 2014 eclipse (see Sect \ref{neowisesec}). }
	\label{neowise}
\end{figure}

In Figure. \ref{neowise} we present archive post-cryo WISE (NEOWISE, \citealt{2011ApJ...731...53M, 2014ApJ...792...30M}) observations of RW Aur between 2014 and 2015. This dataset consists of four blocks of observations, each consisting of several measurements over the course of 1-2 consecutive nights. The first block has been taken prior to the start of the 2014 dimming while the remaining three blocks are in the faint state. The red datapoints represent the average value of the measurements of a given block. The star shows clear variability on timescales of hours and days in all blocks. After the dimming, the average magnitude in W1 and W2 increases by 0.14$-$0.26 and 0.66$-$1.28 mag respectively. The photometric error is $\sim$\,0.03 mag in W1 and $\sim$\,0.01 mag in W2. The changes in the average values in the WISE bands after the eclipse are comparable to the M and L-band brightenings reported by \cite{2015IBVS.6143....1S}, therefore solidifying the evidence for the mid-IR excess emission increase coinciding with the dimming in the optical. 

\subsection{Spectral analysis}

\begin{figure}
\centering
\includegraphics[scale=0.41]{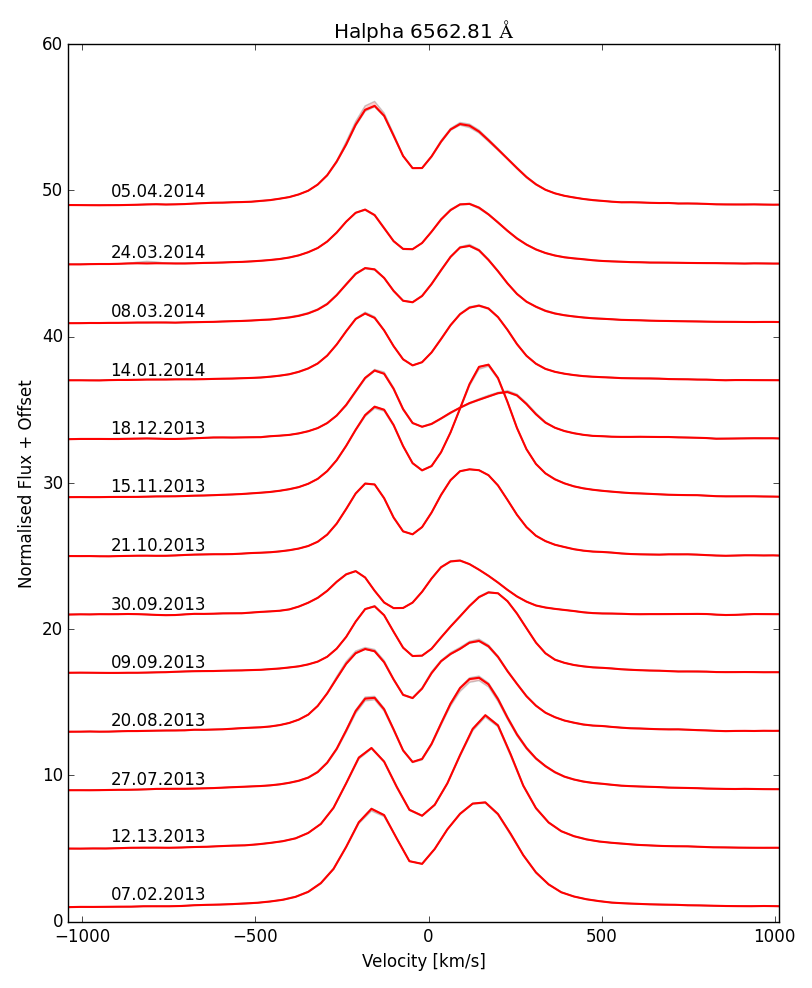} \\
\includegraphics[scale=0.41]{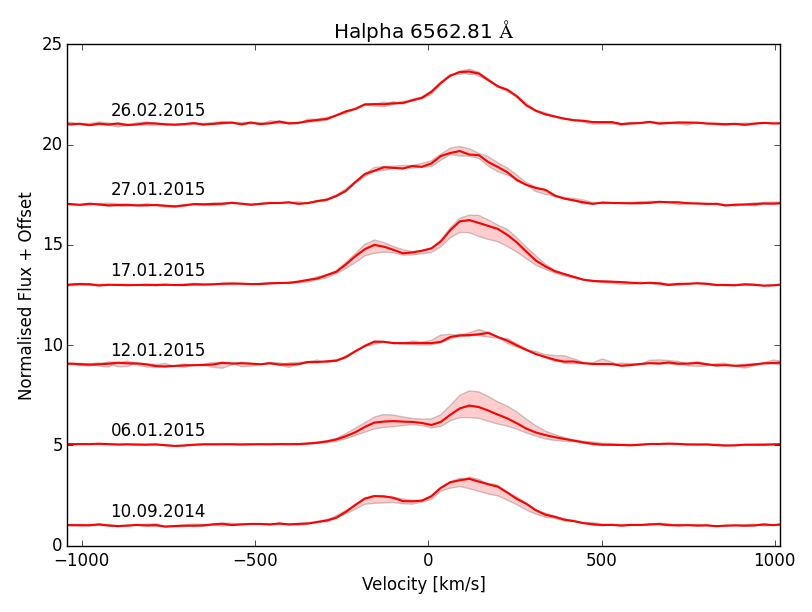}
\caption{Full time-series of all H$\alpha$ profiles. The top panel shows the observations before the dimming event and the bottom panel shows the spectra taken after the start of the eclipse. The red lines indicate the mean profile for that night's observation. The red shading indicates the spread between the individual 3 spectra for that night.   }
\label{HaSeries}
\end{figure}

\begin{figure}
		\includegraphics[scale=0.41]{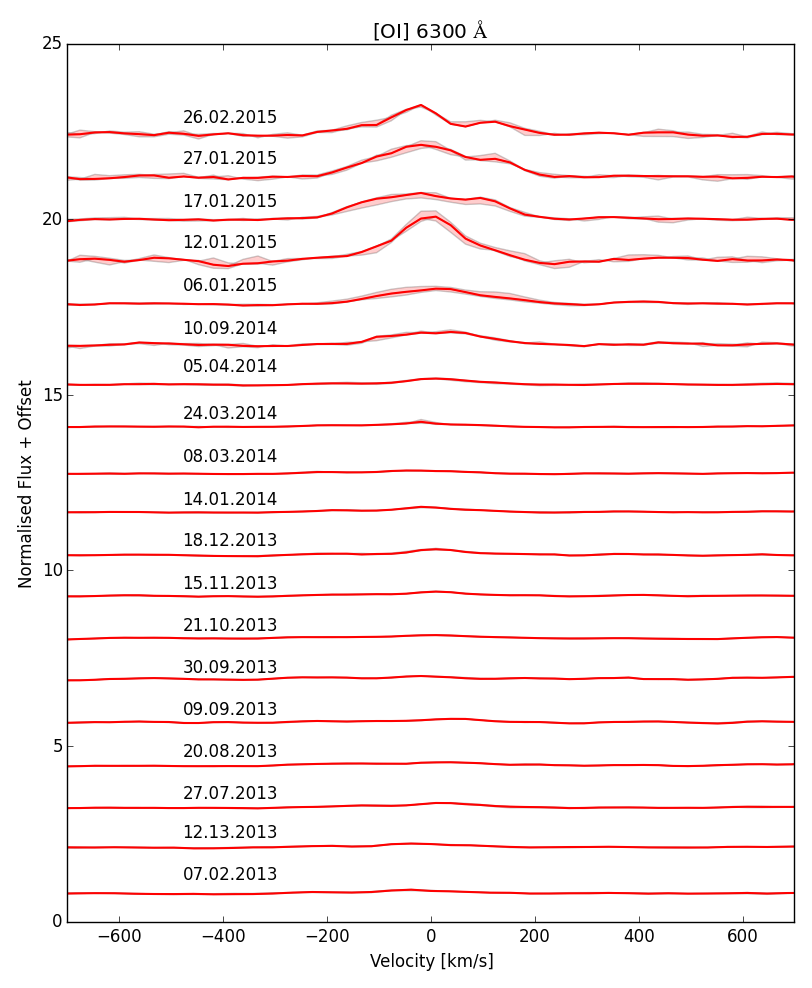} 
\caption{Partial time series of the [O\,I] profiles. The red lines indicate the mean profile for that night's observation. While the red shading indicates the spread between the individual 3 spectra for that night.  }
\label{OISeries}
\end{figure}

\begin{figure}
\includegraphics[scale=0.43]{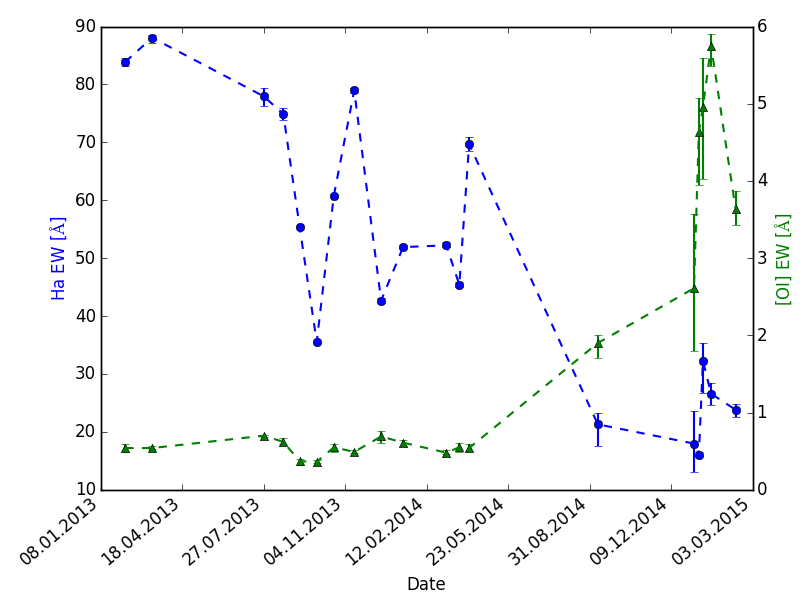}
\caption{ Mean EW versus date for H$\alpha$ (blue dots) and [O\,I] (green triangles). Here the mean EW is taken from the 3 observations each night. The error bars represent the differences between the 3 observations that night. We see an increase in the [O\,I] by a factor of 10 after the dimming. We also observe a drop in the H$\alpha$ EW.}
\label{HA_OI_EW}
\end{figure}

\begin{figure}
\includegraphics[scale=0.41]{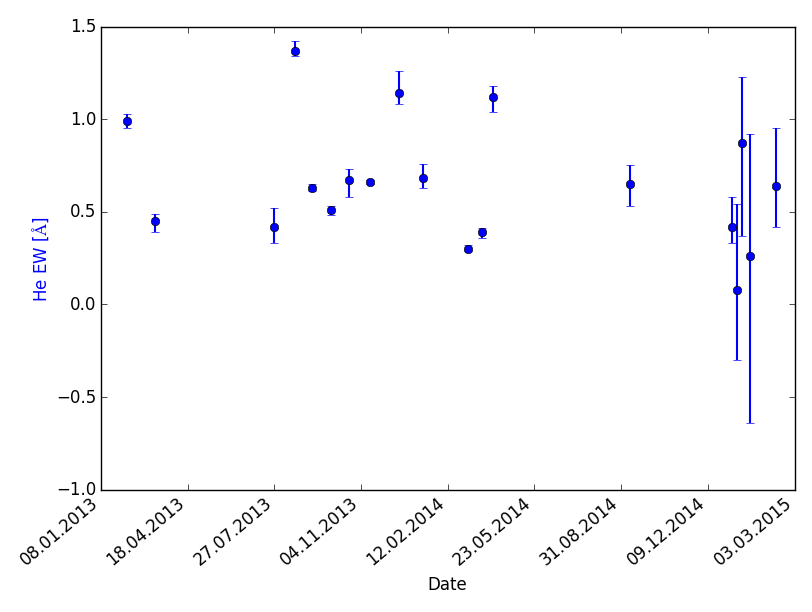}
\caption{Mean EW versus date for He\,I. Here the mean EW is taken from the 3 observations each night. The error bars represent the differences between the 3 observations that night. No significant change in the He\,I EW is observed.}
\label{HeEW}
\end{figure}

The FRODOSpec spectra show clear emission lines typically associated with accretion (H$\alpha$ (6563\,$\AA$) and He I (6678\,$\AA$)) and outflow ([O\,I] (6300\,$\AA$)). We present profile timeseries for the H$\alpha$ and [O\,I] lines in Figure \ref{HaSeries} and \ref{OISeries}. We measure equivalent widths (EWs) for these three lines. There are three measurements for each consecutive night. We normalize all spectra within the same spectral range to minimise any  variability from changing weather conditions. The IFU positioning does not seem to have an effect on EWs. The results are presented in Figure \ref{HA_OI_EW},\,\ref{HeEW} and Table \ref{EWstable}. The errors in the plots are taken to be the spread in measurements over one night (3 observations). We observe a drop in H$\alpha$ and increase in the [O\,I] during the dimming event. The He I shows no significant change. The H$\alpha$ EW drops from 80$-$90$\,\AA$ in the bright state down to $\sim$20\,$\AA$ in the dim state. However, we note the H$\alpha$ EW begins to decrease prior to the 2014 eclipse (Fig. \ref{HA_OI_EW}) and it is therefore possible the change in H$\alpha$ EW is not solely related to the dimming event, as the line is known to be intrinsically variable \citep{2001A&A...369..993P,2005A&A...440..595A}. The [O\,I] EW increases from 0.5 to 5 $\AA$, i.e. by a factor of 10. 


We also show timeseries of the H$\alpha$ and [O\,I] profiles in  Figure \ref{HaSeries} and \ref{OISeries}. In each case the date is given beside the mean profile for that night. The characteristic broad, double-peaked profile of the H$\alpha$ line can clearly be seen in both the bright and dim state. However, during the dimming the entire profile is suppressed and the blue-shifted peak appears significantly weaker than the red-shifted peak. We measure the H$\alpha$ EW for v$<$0\,km\slash s and v$>$0\,km\slash s in order to quantify the difference between the change in the blue and red-shifted peaks after the dimming. The line EW decreases by a factor of $\sim$7 after the eclipse for v$<$0\,km\slash s and by a factor of $\sim$3 for v$>$0\,km\slash s. The difference in the two factors suggests that the blue-shifted peak suppression is indeed significant. In contrast to the H$\alpha$ line, the [O\,I] profile becomes more prominent during the dimming, showing a broad and asymmetric emission.



\section{Discussion}

We present new spectroscopic and photometric observations of RW Aur during the dimming in 2014-2016. In addition, we review the previous and new observational constraints for the system. 

\subsection{Photometric variations}

The dimming event that started in 2014 is continuing through our observing period. The system has become even fainter in the optical during the 2015-16 winter observing season (see Figures \ref{aavso} and \ref{LCOGTlc}).  
In addition to the long-term dimming trend, we clearly see irregular short timescale variations in the JGT (Figure \ref{JGTlc}) and LCOGT (Figure \ref{LCOGTlc}) data.
These variations could be related to small scale changes in extinction along the line of sight. However, the irregular components in the morphology of the JGT lightcurves better resemble burst-like events. Such morphology is typically associated with changes in an unstable accretion flow for T Tau stars (see \citealt{1994AJ....108.1906H,2014AJ....147...82C} for an overview of the different types of photometric variability displayed by T Tau stars). Being able to see such short time-scale irregular variations in RW Aur implies that the accretion flow in the system remains at least partially visible during the dimming.

\begin{figure}

	\includegraphics[scale=0.44]{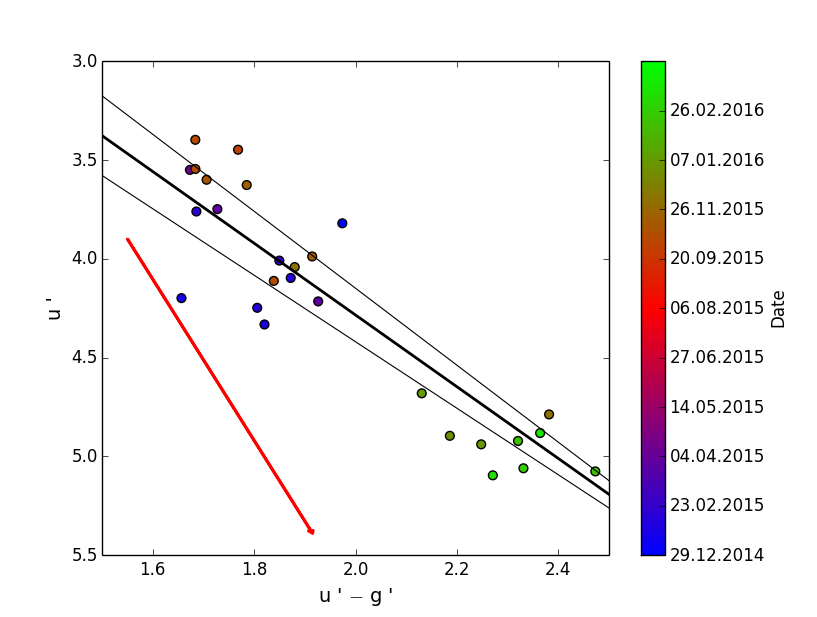}
	 \includegraphics[scale=0.44]{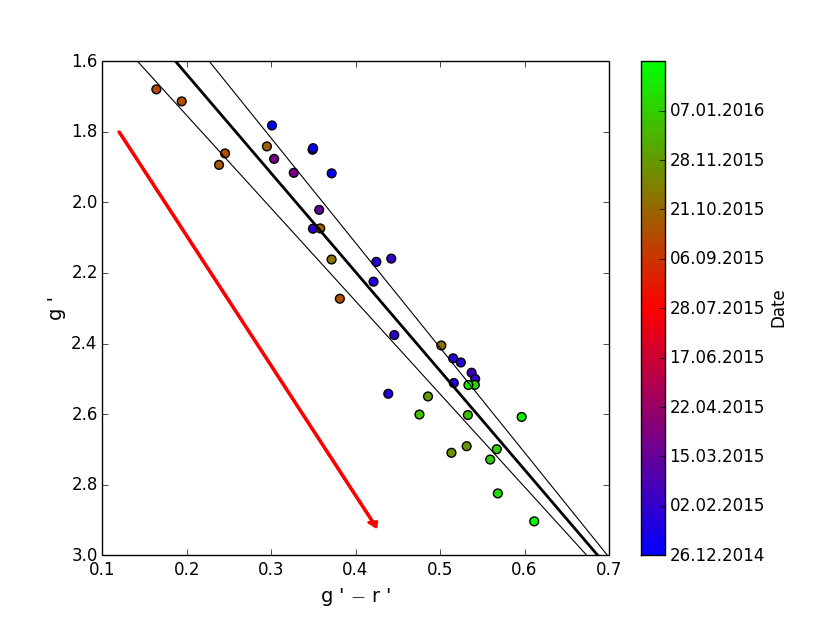}
	\includegraphics[scale=0.44]{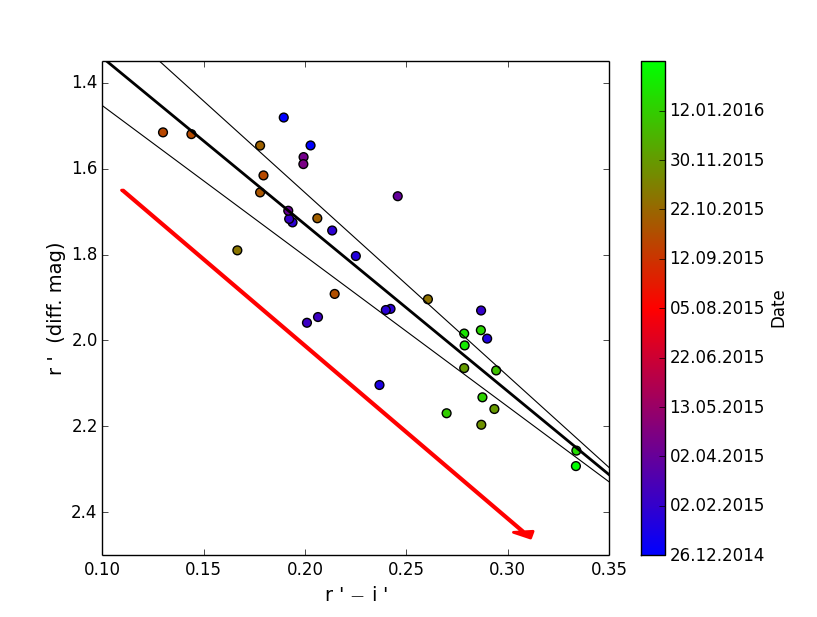} \\

	\caption{ Colour-magnitude diagrams constructed from all LCOGT observations. The epochs of observations are colour coded. The black solid line indicates the best linear fit to the data with thinner lines indicating the error in the slope estimation. The red arrow is the interstellar extinction slope for $R_V = 3.1$ and $A_V = 0.9$, where the $A_V$ value is selected such that the resulting extinction in the r-band best matches the observed amplitude in the r-band lightcurve.} 
	\label{CMDplots}

\end{figure}

In addition to the overall drop in photometric flux in the optical, colour-magnitude diagrams of our LCOGT data (Figure \ref{CMDplots}) reveal a significant long-term reddening trend since the start of the eclipse. We compare the observed reddening with models for interstellar extinction by \cite{1989ApJ...345..245C}. Extinction fails to fit the observed colours simultaneously. In  particular, the model under-predicts the observed u\,\arcmin-band amplitude and is unable to reproduce the u\,\arcmin$-$\,g\,\arcmin reddening slope. We therefore conclude that the long-term reddening trend is not consistent with interstellar-like extinction.  

We also attempt to explain the reddening with a simple hot spot model, where the star and spots are approximated by black bodies and the spot temperature and filling factor are varied in order to fit the observed lightcurve amplitudes (eg. \citealt{2009MNRAS.398..873S}, \citealt{1995A&A...299...89B}). 

We find a number of parameter values that produce amplitudes and reddening slopes similar to the observed ones, especially in the red portion of the spectrum. However, hot spots have one major problem with explaining the observed long-term reddening. Hot spots form at the stellar surface as a result from accreting material near-free falling onto the star and creating shocks. As the star rotates the spots come in and out of view, resulting in a photometric variably on time-scales comparable with the stellar rotational period (typically several days for T Tau stars). This is in contrast to the long-term dimming of RW Aur which has lasted for months. While accretion may still play a role for the short-time scale variability of RW Aur, hot spots cannot be the cause for the observed long-term reddening, where the system has continuously been getting redder for months. 

An alternative explanation for the observed long-term reddening arises from the fact that our photometry does not resolve RW Aur A and B. \cite{2015IBVS.6126....1A} present resolved photometry of the system. Their measurements (after the start of the dimming) show an R-band magnitude of 13.08 and 11.97 for the A and B-components respectively. Our JGT photometry yields R$\sim$12 mag suggesting that the red part of the spectrum in our observations is most likely dominated by the B-component. We can therefore interpret the observed long-term reddening as the A-component becoming faint enough that the B-component becomes the dominant source of flux, making the entire system appear redder while simultaneously getting dimmer as we lose flux from RW Aur A. 

\subsection{Spectral signatures}

Spectroscopy provides more information about accretion and outflow in young stellar systems. In the classical magnetospheric accretion scenario the inner-most region of the circumstellar disk is threaded by stellar magnetic field lines. Material is channelled from the disk onto the star forming accretion columns at near free-fall velocities, with a shock forming at the surface of the star. The shock and accretion columns generate H$\alpha$ emission. The accretion-related H$\alpha$ emission lines are typically broad with equivalent widths of over 10\,$\AA$. Other emission lines related to CTTS accretion include permitted lines of He\,I, O\,I and Ca\,II. For a review of accretion signatures we refer the reader to \cite{2005ApJ...626..498M} and references therein. In addition to accretion indicators, CTTS often shows outflow-related forbidden line emission such as [O\,I] lines \citep{1995ApJ...452..736H,1984A&A...141..108A}.   

Our spectra show that the EW of the [O I] (6300\,$\AA$) line increases by a factor of 10 after the eclipse (Figure \ref{OISeries},\ref{HA_OI_EW}). The observed EW increase is in agreement with measurements by \cite{2015A&A...577A..73P} and \cite{2016ApJ...820..139T}. \cite{2015A&A...577A..73P} attribute the change in the forbidden line's EW
to a drop in the photospheric continuum, as reported by the resolved photometry of RW Aur A by \cite{2015IBVS.6126....1A}. \cite{2015A&A...577A..73P} further imply that the obscuring body must be blocking the inner parts of the system while the outflow remains visible. 

The He I (6678\,$\AA$) and H$\alpha$ emission lines remain clearly visible in our spectra after RW Aur enters the eclipse. This again indicates that at least part of the accretion flow is still visible. The He I EW remains constant (Figure \ref{HeEW}), implying the He-emitting regions are obscured in the same way as the photosphere. 

The H$\alpha$ EW decreases after the eclipse (Figure \ref{HA_OI_EW}). Our EW measurements show lower values for the line during the dimming compared to the values reported by both \cite{2015A&A...577A..73P} and \cite{2016ApJ...820..139T}. We attribute this to the fact that our spectroscopy does not resolve the A and B-components, hence we get an additional flux in the continuum from RW Aur B, resulting in a lower EW value.    

The H$\alpha$ line profiles show a suppressed blue-shifted peak after the start of the dimming (Figure \ref{HaSeries}). 
This can also be seen in the H$\alpha$ profiles from \cite{2016ApJ...820..139T} from both the 2010 and 2014 dimmings.
This blue-shifted absorption can be interpreted as the presence of an absorbing material moving towards the observer's line-of-sight, i.e. an outflow.

\subsection{Origin of the eclipse}

In Sect. 1 we introduce two possible explanations for the long dimming event observed in RW Aur. The obscuring screen could either be a tidal arm in the outer disk, as suggested by \cite{2013AJ....146..112R,2016AJ....151...29R},
or a wind emanating in the inner disk, suggested by \cite{2015A&A...577A..73P}.
Our observations provide new evidence to support the occultation by a wind. 

In particular, we show that the blue-shifted portion of the H$\alpha$ line is suppressed since the beginning of the eclipse, indicating the presence of outflowing and thus blueshifted gas with velocities of 100-200\,kms$^{-1}$ in the occulting screen. This scenario is further supported by the increase in the mid-IR fluxes of the system in eclipse, reported by \cite{2015IBVS.6143....1S} and visible in archive NEOWISE data in the W2 band (Figure. \ref{neowise}). \cite{2015IBVS.6143....1S} fit a blackbody to the spectral energy distribution of RW Aur in its faint state and find that the excess IR emission is plausibly explained by the presence of hot dust with temperature around 1000\,K along the line of sight.


Thus, the occulting body is outflowing and contains hot dust, which fits with a scenario where a wind emanating from the inner disk eclipses the star RW Aur A. These two facts together are also inconsistent with the tidal arm scenario, in which the occulting material would not be expected to move rapidly along the line of sight and the dust in the screen would be cold. If the eclipse indeed arises in a wind, the long duration of the current event would imply that the outflowing screen stretches azimuthally over a large portion of the inner disk. We note that the recent resolved X-ray detection of RW Aur shows a slight soft X-ray extension of RW Aur A along the blue-shifted jet axis close to the star  \citep{2014ApJ...788..101S} - this could be related to the obscuring screen discussed here. Furthermore, \cite{2015A&A...584L...9S} present resolved X-ray and NIR observations of the system in eclipse and provide further evidence
for the presence of an obscuring screen with hot material located close to the star. 

In addition, there is also a geometrical argument in favour of a blocking screen located above the inner disk. The CO map modelling from \cite{2006A&A...452..897C} found an inclination between 45-60$^{\circ}$ for the large-scale disk. The inclination of the jet to the line of sight is measured to be $46\pm 3^{\circ}$ \citep{2003A&A...405L...1L}. These values indicate that the large-scale disk is far from edge-on, which makes it difficult to lift material in the outer disk into the line of sight. In fact, the hydrodynamical simulations of the tidal arm scenario by \cite{2015MNRAS.449.1996D} give a best fitting value of 64$^{\circ}$ for the inclination, and thus does not match the constraints from observations.

On the other hand, the modelling of K-band Keck interferometric observations tracing the continuum and gas in the inner region of the disk results in an inclination angle of 75$^{\circ}$ (\citealt{2014MNRAS.443.1916E}, Table 6) for an emitting region within a fraction of 1\,AU. This is a significantly higher inclination and closer to edge-on than for the outer disk. We note that some of the Keck interferometer data for RW Aur A was taken during the first eclipse in 2010/11. The discrepancy in the inclinations between inner and outer disk may be evidence for a large-scale warp in the disk, possibly caused by interaction with the companion RW Aur B. Thus, RW Aur B may have an effect on the disk, but it is unlikely to be the direct cause of the feature that eclipses RW Aur A. 

Adopting the value from the inner disk for the disk inclination, the material in the occulting screen needs to be vertically displaced from the disk midplane by at least 0.26 times the distance from the central object. This value is comparable to the maximum vertical extent of the inner disk found in the so-called $``$dippers$"$, T Tauri stars with periodic eclipses caused by disk warps caused by accretion channelled by a magnetic field tilted with respect to the rotational axis \citep{1999A&A...349..619B,2015A&A...577A..11M}. RW Aur is not known to be a dipper, perhaps because the tilt of its magnetic field is not sufficient. Based on the high value for the inner disk inclination, however, it is conceivable that a magnetically driven disk wind can lift material into the line of sight.

The jet of RW Aur A has been extensively studied with high-resolution imaging. It is one of the densest among the T Tauri stars investigated in detail, but the mass outflow rate seems to be stable within an order of magnitude along the jet axis and in both lobes at $10^{-9}$ to $10^{-8}\,M_{\odot}$yr$^{-1}$ \citep{2009A&A...506..763M}. However, in the past there have been clear indications for strong variability in the forbidden lines caused by the wind as well as in the blue-shifted portion of H$\alpha$ \citep{2005A&A...440..595A}.

In the standard scenario of magnetospheric accretion, the wind is thought to be coupled to the accretion, thus, we would expect that a change in the outflow rate or geometry, as indicated by our interpretation of the RW Aur dimming, is preceded by strong variability in the accretion flow. As far as we are aware, there is no evidence for a recent accretion burst in RW Aur. Its accretion rate is relatively high ($10^{-7}$ to $10^{-6}\,M_{\odot}$yr$^{-1}$) compared with other CTTS with similar mass, but seems to be relatively stable (e.g., \citealt{2014MNRAS.440.3444C}). Thus, changes in the wind characteristics might not always be coupled to changes in the accretion properties. One possible trigger for a wind outburst could be an interaction between the disk and the putative stellar or substellar spectroscopic companion to RW Aur A \citep{1999A&A...352L..95G,2001A&A...369..993P}. However, it is unclear how such an interaction can produce the sudden onset of wind activity multiple times over the course of the last $\sim$\,7 years, with no precedents observed in the previous $\sim$\,50 years. Another alternative may be that the fly-by of RW Aur B has caused an instability in the disk of RW Aur A, although modelling this scenario is beyond the scope of this study.


We encourage further multi-band photometric and spectroscopic monitoring of accretion and outflow signatures of RW Aur to further test and solidify our understanding of the system.

\section{Acknowledgements}

We acknowledge with thanks the variable star observations from the AAVSO International Database contributed by observers worldwide and used in this research. This publication also makes use of data products from NEOWISE, which is a project of the Jet Propulsion Laboratory/California Institute of Technology, funded by the Planetary Science Division of the National Aeronautics and Space Administration. The Liverpool Telescope is operated on the island of La Palma by Liverpool John Moores University in the Spanish Observatorio del Roque de los Muchachos of the Instituto de Astrofisica de Canarias with financial support from the UK Science and Technology Facilities Council. This work makes use of observations from the LCOGT network, taken under programs STA2014B-002 and STA2015A-002. The authors acknowledge support from the Science \& Technology Facilities Council through grants no.ST/K502339/1 and ST/M001296/1 and the Science Foundation Ireland through grant no.10/RFP/AST2780.

\bibliographystyle{mnras}
\bibliography{inb}






\end{document}